\documentclass[11pt]{article}

\pdfoutput=1

\usepackage{cite}
\usepackage{hyperref}
\usepackage{cancel}

\usepackage{lineno}

\usepackage{amsmath}
\usepackage{amssymb,amsfonts,amsthm}
\usepackage{epsfig}
\usepackage{graphicx}
\usepackage{caption}
\usepackage{subcaption}

\usepackage{url} 
\usepackage{bm} 
\hypersetup{colorlinks=true}
\usepackage{slashed}
\usepackage{cite}

\usepackage{color}

\usepackage{rotating}

\usepackage{fancyhdr}
\usepackage{textcomp}

\usepackage{float} 
\usepackage[utf8]{inputenc}

\newcommand{\unit}{\leavevmode\hbox{\small1\kern-3.6pt\normalsize1}}

\newcommand{\gsim}{\lower.7ex\hbox{$\;\stackrel{\textstyle>}{\sim}\;$}}
\newcommand{\lsim}{\lower.7ex\hbox{$\;\stackrel{\textstyle<}{\sim}\;$}}

\newcommand{\be}{\begin{equation}}
\newcommand{\ee}{\end{equation}}

\newcommand{\bea}{\begin{eqnarray}}
\newcommand{\eea}{\end{eqnarray}}

\def\simgt{\stackrel{>}{{}_\sim}}

\def\sigsi{\sigma^{SI}_{S p}}

\begin{document}

\title{Extended Higgs-portal dark matter and the {\it Fermi}-LAT Galactic Center Excess}

\thispagestyle{empty}
\begin{flushright}

 {\small 
 IFT-UAM/CSIC-17-113
}\\

\end{flushright}

\begin{center}
  {\bf {\LARGE Extended Higgs-portal dark matter and \\[1mm] the {\it Fermi}-LAT Galactic Center Excess}}

\renewcommand*{\thefootnote}{\fnsymbol{footnote}}

  \vspace{0.5cm}
  {\large
  J.~A.~Casas $^{a}$,
    G.~A.~G\'omez~Vargas $^{b}$,
    J.~M.~Moreno $^{a}$, 
    J. ~Quilis $^{a}$ \\
   and
    R.~Ruiz de Austri$^{c}$
  }
  \\[0.2cm] 

  {\footnotesize{
       $^a$ Instituto de F\'{\i}sica Te\'{o}rica UAM/CSIC, Universidad Aut\'{o}noma de Madrid,  28049, Madrid, Spain\\
       $^b$ Instituto de Astrof\'isica, Pontificia Universidad Cat\'olica de Chile, Avenida Vicu\~na Mackenna
4860, Santiago, Chile\\
      $^c$ Instituto de F\'{i}sica Corpuscular, IFIC-UV/CSIC, Valencia, Spain\\
        }
    }

\vspace*{0.7cm}

  \begin{abstract}
In the present work, we show that the Galactic Center Excess (GCE) emission, as recently updated by the {\it Fermi}-LAT Collaboration, could be explained by a mixture of Fermi-bubbles-like emission plus dark matter (DM) annihilation, in the context of a scalar-singlet Higgs portal scenario (SHP). In fact, the standard SHP, where the DM particle, $S$, only has renormalizable interactions with the Higgs, is non-operational due to strong constraints, especially from DM direct detection limits.
Thus we consider the most economical extension, called ESHP (for extended SHP), which consists solely in the addition of a second (more massive) scalar singlet in the dark sector. The second scalar can be integrated-out, leaving a standard SHP plus a dimension-6 operator. Mainly, this model has only two relevant parameters (the DM mass and the coupling of the dim-6 operator). DM annihilation occurs mainly into two Higgs bosons, $SS\rightarrow hh$. We demonstrate that, despite its economy, the ESHP model provides an excellent fit to the GCE (with p-value $\sim 0.6-0.7$) for very reasonable values of the parameters, in particular, $m_S \simeq 130$ GeV. This agreement of the DM candidate to the GCE properties does not clash with other observables and keep the $S-$particle relic density at the accepted value for the DM content in the universe.
   \end{abstract}
\end{center}

\newpage
\pagestyle{plain}

\section{Introduction}
\label{sec:introduction}

\begin{figure}[t!]
\centering 
\includegraphics[width=0.8\linewidth]{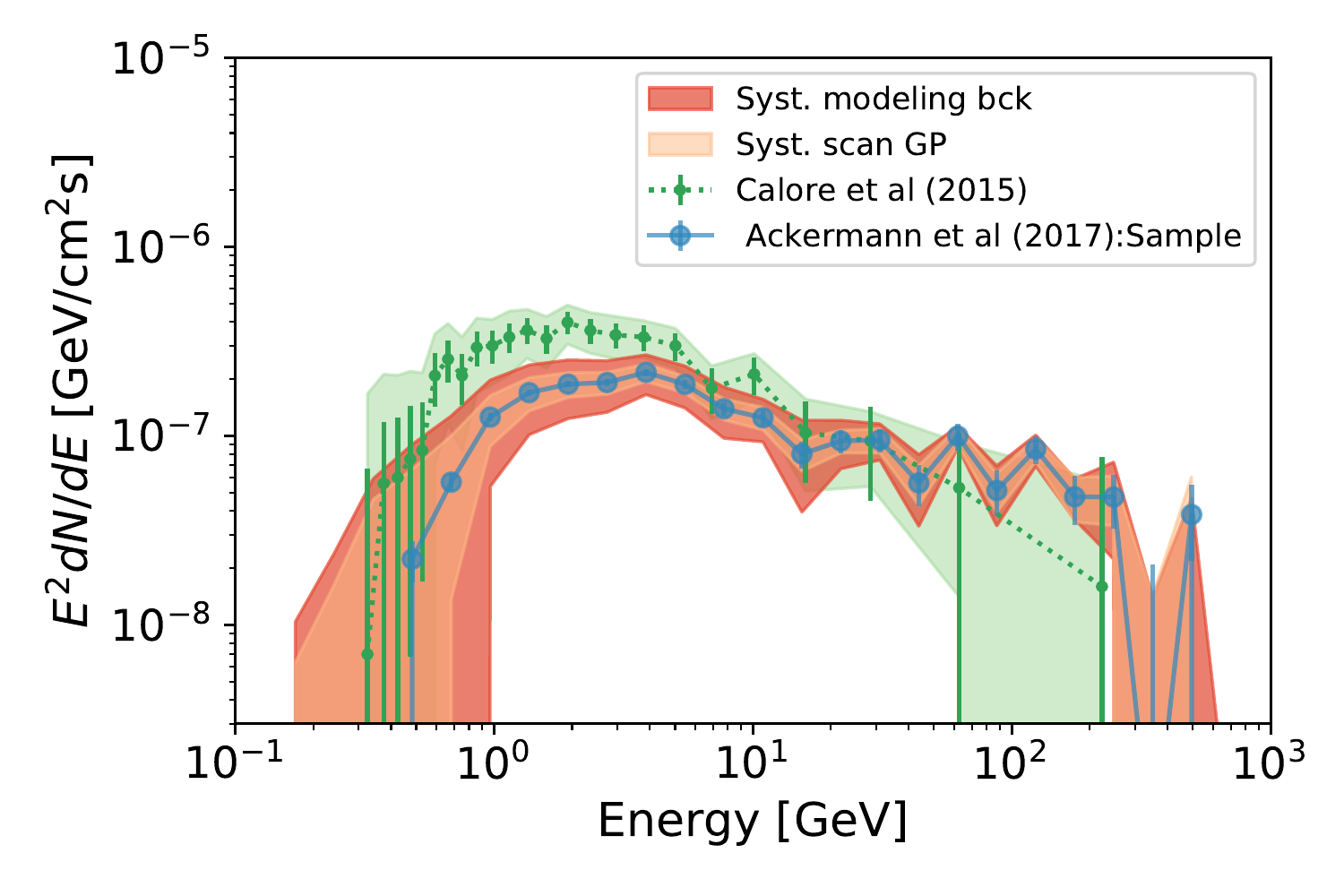}
\caption{ Blue points: GCE spectrum from ref.~\cite{1704.03910} using the Sample model, see the reference for more info. The light orange band represents the diagonal of the covariance matrix due to excesses along the Galactic Plane,
 obtained using the same procedure as for the GCE~\cite{Achterberg:2017emt}. The dark orange band is the diagonal of the covariance matrix from variations in the GCE due to uncertainties in modelling diffuse emission from ref.~\cite{1704.03910}.
For comparison  we include the green points and shadow green area, obtained in the analysis of ref.~\cite{Calore:2014xka}.  
}
\label{fig:GCE}
\end{figure}

The Large Area Telescope (LAT)~\cite{2009ApJ...697.1071A} onboard the {\it Fermi} satellite has revealed the $\gamma$-ray sky with unprecedented detail, prompting the study of models of fundamental physics beyond the Standard Model (SM) in different ways~\cite{Meyer:2016wrm,Vasileiou:2013vra,Charles:2016pgz}. 
Diverse studies of the {\it Fermi}-LAT data have identified that the Galactic Center area is brighter than expected from conventional models of diffuse $\gamma$-ray emission. 
Including in the interstellar diffuse emission fitting procedure a template compatible with predictions of DM annihilating to SM particles, following a slightly contracted Navarro-Frenk-White (NFW) profile, substantially improves the description of the data. The emission assigned to this extra template is the so-called Galactic Center excess (GCE)~\cite{Goodenough:2009gk, Vitale:2009hr, Hooper:2010mq, Gordon:2013vta, Hooper:2011ti, Daylan:2014rsa, 2011PhLB..705..165B, Calore:2014xka, Zhou:2014lva,TheFermi-LAT:2015kwa, 1704.03910}. 
In the recent analysis of ref.~\cite{1704.03910}, to study the GCE a representative set of models from ref.~\cite{Ackermann:2012pya} was taken into account, along with different lists of detected point sources. The bottom line is that there remains a GCE emission, now peaked at the $\sim 3$ GeV region, i.e. slightly shifted towards higher energies with respect to previous studies. In this paper we will consider the GCE emission obtained by using the so-called Sample Model (light blue points in Fig.~\ref{fig:GCE}, see sect. 2.2 of ref.~\cite{1704.03910} for details on the model) and a combination of the covariance matrices derived in ref.~\cite{Achterberg:2017emt} in order to perform the fits.

The nature of the GCE is under debate. Apart from the DM hypothesis, it has been proposed that the GCE can be due to collective emission of a population of point sources too dim to be detected individually~\cite{McCann:2014dea,Mirabal:2013rba,2015arXiv150402477O,2014JHEAp...3....1Y,Petrovic:2014xra,Cholis:2014lta,Macias:2016nev},  or the result of fresh cosmic-ray particles injected in the Galactic Center region interacting with the ambient gas or radiation fields, see for instance ref.~\cite{Gaggero:2015nsa,Porter:2017vaa}. 
Indeed, some studies favour a point-source population as explanation to the GCE emission~\cite{2016PhRvL.116e1103L,2016PhRvL.116e1102B,Storm:2017arh,Fermi-LAT:2017yoi,Eckner:2017oul,Bartels:2017vsx}, however further investigations on the data are required, since the GCE could be the result of a combination of phenomena at work in the inner Galaxy, including DM annihilation~\cite{Caron:2017udl}.  
On the other hand, the GCE may well have different origins below and above $\sim 10$ GeV~\cite{Linden:2016rcf,1704.03910}. The high energy tail ($E>10$ GeV) could be due to the extension of the {\it Fermi} bubbles observed at higher latitudes~\cite{1704.03910} or some mismodeling of the interstellar radiation fields and a putative high-energy electron population~\cite{Porter:2017vaa}. At lower energies ($E<10$ GeV) the GCE might be due to DM annihilation, unresolved millisecond pulsars (MSP), or a combination of both, see for instance refs.~\cite{Caron:2017udl,Achterberg:2017emt}. According to ref.~\cite{1704.03910}, the interpretation of the GCE as a signal for DM annihilation is not robust, 
but is not excluded either. Actually, it has been claimed in refs.~\cite{Hooper:2015jlu,Hooper:2016rap,Haggard:2017lyq} that a population of $\gamma$-ray pulsars cannot be responsible for the entire GCE emission. 

On the other side, the interpretation of the GCE as originated by DM (with or without an astrophysical source for the high energy tail) is not straightforward, particularly in theoretically-sound models (see, for instance, Refs.~\cite{Modak:2013jya,
Okada:2013bna,
Banik:2014eda,
Balazs:2014jla,
Agrawal:2014oha,
Ghorbani:2014gka,
Hooper:2014fda, 
Martin:2014sxa,
Ko:2014loa, 
Basak:2014sza, 
Wang:2014elb, 
Abdullah:2014lla,
Cerdeno:2015ega,
Cline:2015qha,
Duerr:2015bea,
Kim:2015fpa,
Chen:2015nea,
Kim:2016csm,
Horiuchi:2016tqw,
Escudero:2016kpw,
DuttaBanik:2016jzv,
Tunney:2017yfp,
Escudero:2017yia}).
The most common difficulty is that a DM model able to reproduce the GCE also leads to predictions on DM direct detection which are already excluded by present experiments.  The PICO-60~\cite{Amole:2017dex}  (spin dependent cross-section) and  XENON1T~\cite{Aprile:2017iyp} (spin-independent cross-section) experiments provide the most stringent direct detection bounds, so far. 
Of course, things change for better if only a fraction of the low-energy GCE is associated to DM annihilation. For example, 
in the recent paper~\cite{Achterberg:2017emt} it is shown that supersymmetric DM could be well responsible of $\sim 40\%$ of the low-energy ($E<10$ GeV) GCE emission. However, fitting the whole low-energy GCE just with DM emission is more challenging.

One of the most economical scenarios for DM is
the Scalar Singlet-Higgs-Portal (SHP) model~\cite{Silveira:1985rk,McDonald:1993ex,Burgess:2000yq}
 where the DM particle is a singlet scalar, $S$, which interacts with the SM matter through couplings with the Higgs field, $H$.
The relevant Lagrangian of the simplest SHP model contains only renormalizable terms and reads 
\begin{equation}
\mathcal{L}_{\rm SHP}=\mathcal{L}_{\rm SM}+\frac{1}{2}\partial_{\mu} S \partial^{\mu} S- \frac{1}{2}m_0^2 S^2-\frac{1}{2}\lambda_S |H|^2 S^2 -\frac{1}{4}\lambda_{4} S^4 .
\label{SHPlagr}
\end{equation}
Here it is assumed, for simplicity, that $S$ is a real field (the modification for the complex case is trivial), subjected to a discrete symmetry $S\rightarrow -S$ in order to ensure the stability of the DM particles; otherwise, the above renormalizable Lagrangian is completely general. After electroweak (EW) breaking, the neutral Higgs field gets a vacuum expectation value, $H^0=(v+h)/\sqrt{2}$, giving rise to new terms, in particular a trilinear coupling between $S$ and the Higgs boson, $(\lambda_S v/2) h S^2$. 
Since the DM self-coupling, $\lambda_4$, plays little role in the DM dynamics, the model is essentially determined by two parameters: $\{m_0, \lambda_S\}$ or, equivalently, $\{m_S, \lambda_S\}$, where $m_S^2=m_0^2+\lambda_Sv^2/2$ is the physical $S-$mass after EW breaking. Consequently, requiring that the observed DM relic density is entirely made of $S-$particles essentially corresponds to a line in the $\{m_S, \lambda_S\}$ plane \cite{Cline:2013gha,Feng:2014vea}. This line becomes a region if one only requires the present density of $S-$particles to be a component of the whole DM relic density, i.e. $\Omega_S\leq \Omega_{CDM}$.

The SHP model is subject to important  observational constraints (in particular, bounds from DM direct-detection) which rule out large regions of the parameter space, see e.g. ref.\cite{Casas:2017jjg} for a recent update. 
(The allowed region is somewhat larger when $\Omega_S\leq \Omega_{CDM}$ is tolerated.) As a consequence,  only a narrow range of masses around the resonant region ($m_S\sim m_h/2$) plus the region of higher masses ($m_S \simgt 500$ GeV) survive. 
 When one tries to fit the (low-energy) GCE with the SHP, it turns out that only the 
 (fairly tuned)
 resonant region can give some contribution  
 to the flux excess, 
 but still much lower than needed  \cite{Cuoco:2016jqt}. The optimal result 
 (maximum contribution)
 is achieved 
  when $\lambda_S$ is such that  $\Omega_S=\Omega_{CDM}$, i.e. when
  the S-relic-density is maximal. 
In this paper, we re-visit this issue by considering a particularly economical extension of the SHP model, the so-called Extended-SHP (ESHP) model, which simply consists in the addition of a second (heavier) scalar singlet in the dark sector  \cite{Casas:2017jjg}. This extra particle can be integrated-out, leaving a standard SHP plus a dimension-6 operator, $S^2(|H|^2-v^2/2)^2$, which reproduces very well the ESHP results. In consequence, the model essentially adds just one extra parameter to the ordinary SHP. The main virtue of the ESHP is that it allows to rescue large regions of the SHP, leading to the correct relic density and avoiding the strong DM direct-detection constraints. 

In section 2 we introduce the ESHP, explaining the main features of its phenomenology. In section 3 we explain the procedure followed to fit the GCE  with the ESHP model. In section 4 we present the results, showing that, for large regions of the parameter space, the ESHP model provides excellent fits to the GCE. The conclusions are presented in section 5.

\section{Extended Scalar-Higgs-Portal dark matter (ESHP)}

As mentioned in the Introduction, the  ESHP model simply consists in the addition of a second scalar-singlet to the SHP model. Denoting $S_1, S_2$ the two scalar particles, subject to the global $Z_2$ symmetry, $S_1\rightarrow -S_1$, $S_2\rightarrow -S_2$ in order to guarantee the stability of the lightest one, the most general renormalizable Lagrangian reads
\begin{eqnarray}
\mathcal{L}_{\rm ESHP}&=&\mathcal{L}_{\rm SM}+\frac{1}{2}\sum_{i=1,2}\left[(\partial_{\mu} S_i)^2 -m_i^2 S_i^2-\frac{1}{12}\lambda_{i4}S_i^4\right]
\nonumber\\
&&-\frac{1}{2}\lambda_1S_1^2|H|^2-\frac{1}{2}\lambda_2S_2^2|H|^2-\lambda_{12} S_1 S_2 \left(|H|^2 - \frac{v^2}{2}\right)\  +\ \cdots,
\label{ESHP}
\end{eqnarray}
where the dots stand for quartic interaction terms just involving $S_1, S_2$, which have very little impact in the phenomenology.
The terms shown explicitly in the second line are responsible for the DM/SM interactions in the Higgs sector. Once more, after EW breaking, $H^0 = (v+h)/\sqrt{2}$, there appear new terms, such as $(\lambda_{12} v)\, h S_1S_2$.
We have chosen the definition of $S_1, S_2$, so that they correspond to the mass eigenstates with masses $m_{S_i}^2 = m_i^2 + \lambda_i v^2/2$; hence the form of the last term in eq.(\ref{ESHP}). 
We denote $S_1$ the lightest mass eigenstate of the dark sector, and thus the DM particle.

\begin{figure}[t!]
\centering 
\includegraphics[width=0.7\linewidth]{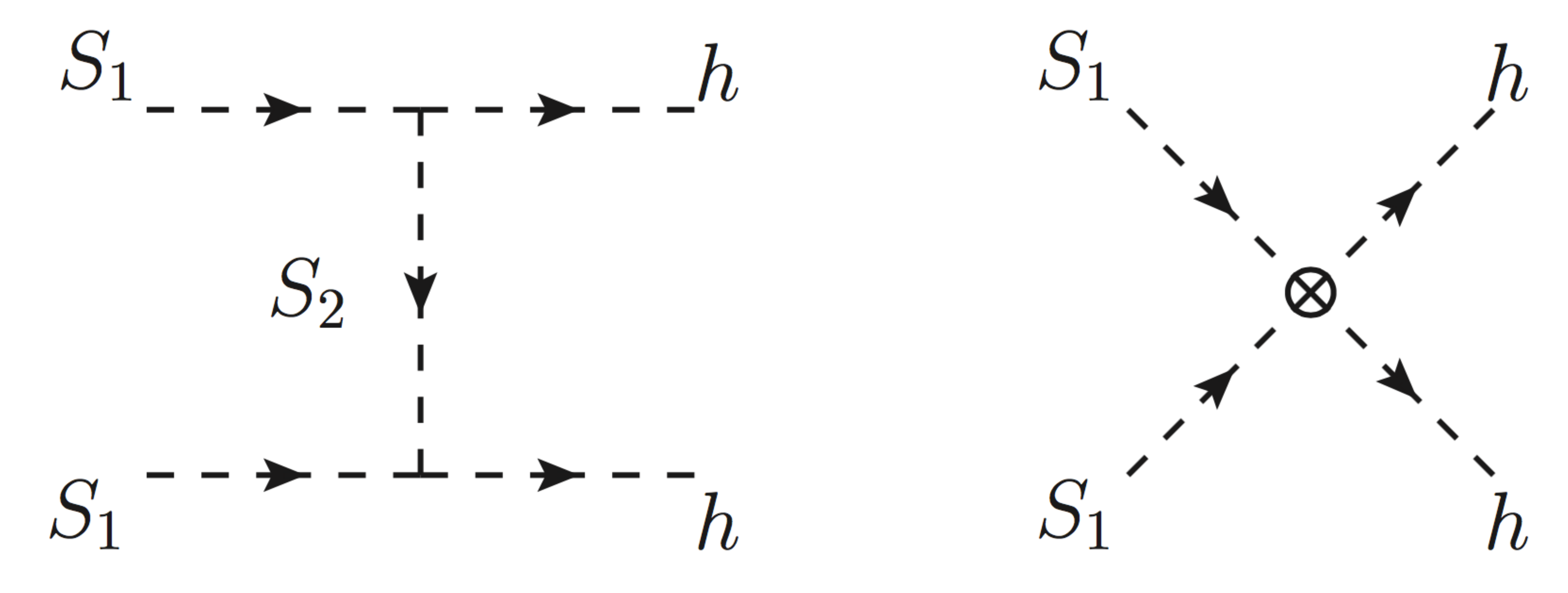}
\caption{
Annihilation $SS\rightarrow hh$ processes of DM in the ESHP model. The  second diagram denotes the same process in the effective theory description.
}
\label{fig:processes}
\end{figure}

The presence of the extra dark particle ($S_2$) leads to new annihilation and/or co-annihilation channels, making it possible to reproduce the correct relic abundance, even if the usual interaction of the DM particle ($S_1$) with the Higgs, $\lambda_1$, is arbitrarily small \cite{Casas:2017jjg}.
This allows to easily avoid the bounds from direct and indirect DM searches. We will focus here in the case where $S_2$ is substantially heavier than $S_1$. Then the main annihilation channel is $S_1S_1\rightarrow h h$, exchanging $S_2$ in $t-$channel, left panel of Fig.~\ref{fig:processes}. 

In this context, $S_2$ can be integrated-out, leading to an effective theory, whose Lagrangian is as in the ordinary scalar-singlet Higgs-portal, eq.(\ref{SHPlagr}), plus additional (higher-dimension) operators\footnote{For notational coherence with the standard SHP, we rename  $S_1\rightarrow S$, $m_{S_1}\rightarrow m_S$, $\lambda_1\rightarrow \lambda_S$.},
\begin{eqnarray}
\mathcal{L}^{\rm eff}_{\rm SHP}=\mathcal{L}_{\rm SHP}
-\frac{1}{2}
\frac{\lambda'} {(500\  {\rm GeV})^2}\ S^2 \left(|H|^2 - \frac{v^2}{2}\right)^2 +\ \cdots ,
\label{Leff}
\end{eqnarray}
where the dots stand for higher-order terms in $S$ or $H$, and $\lambda' = \lambda_{12}^2 (500\ {\rm GeV}/m_{S_2})^2$. Fig.~\ref{fig:processes}, right panel, shows the DM annihilation process in the effective theory language, through the new operator.

A crucial feature of the above higher-order operator is that, after EW breaking, it leads to quartic (and higher) couplings between $S$ and the Higgs boson, $S^2(h^2+2vh)^2$, without generating new cubic couplings, $S^2 h$ (as a usual quartic coupling does). This fact is also obvious in the complete theory (see eq.(\ref{ESHP}) and left panel of Fig.~\ref{fig:processes}), since $S_1, S_2$ are defined as mass eigenstates.
Then, keeping the initial quartic coupling, $\frac{1}{2}\lambda_S S^2|H|^2 $, small, one can enhance the $S$ annihilation without contributing to direct-detection processes (or to the Higgs invisible-width).

\begin{figure}[h!]
\centering 
\includegraphics[width=0.7\linewidth]{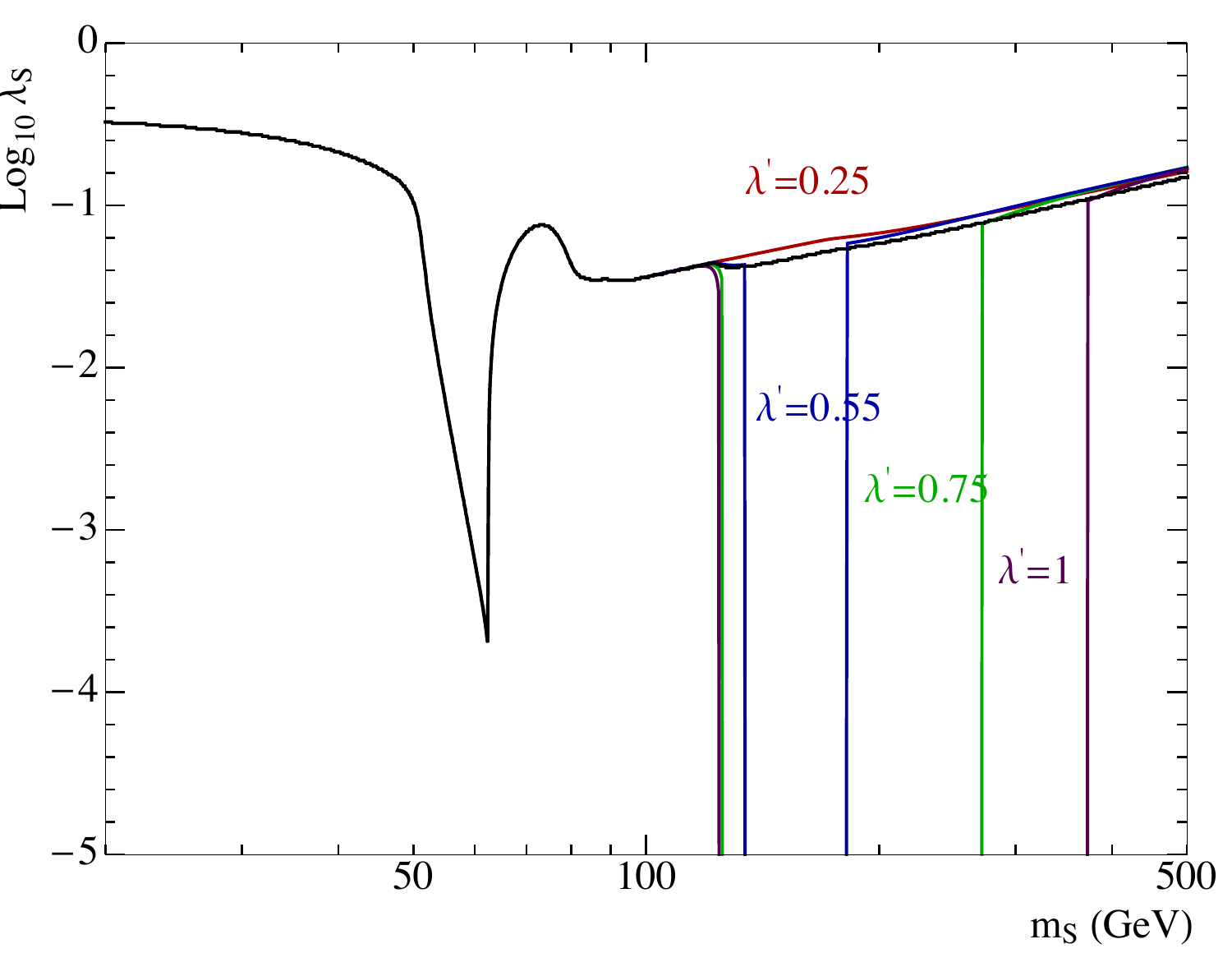}
\caption{
Contour lines of the correct relic DM abundance in the ESHP model for different values of the $\lambda'$ coupling.
}
\label{fig:effective_th}
\end{figure}

Fig.~\ref{fig:effective_th} shows the performance of this effective SHP scenario. The lines shown in the $\{\lambda_S, m_S\}$ plane correspond to the correct relic abundance for different values of the effective coupling $\lambda'$ . Note that
the contribution from the effective operator becomes noticeable for $m_S>m_h/2$, i.e.
when the annihilation channel into a pair of Higgs bosons gets kinematically allowed.
For fixed $m_S$, as $\lambda'$  increases, the value of $\lambda_S$ required to recover the observed relic abundance decreases, quickly becoming irrelevant. 
Then, for any $m_h/2<m_S\lsim 500$ GeV, there is a corresponding value of $\lambda'$  that does the job.

In summary, the effective SHP scenario derived from the ESHP model contains three relevant parameters: $\{\lambda_S, \lambda', m_S\}$. In the next section we will study how well  can this model fit the GCE emission, without conflicting with other observations.

\section{Fitting the Galactic-Center Excess in the ESHP model}

We will assume the GCE is originated by two different sources: an astrophysical one plus the emission from DM annihilation,
\begin{eqnarray}
\Phi_{\rm tot} = \Phi_{\rm astrophysical} + \Phi_{\rm DM}.
\label{flux}
\end{eqnarray}
Concerning the astrophysical component, following the morphological studies in ref.~\cite{1704.03910}, a sensible hypothesis is that it
is a continuation to lower Galactic latitudes of the {\it Fermi} bubbles. Above $10^{\circ}$ in Galactic latitude the spectral shape of the {\it Fermi} bubbles is well characterized by a power law, with index  $1.9 \pm 0.2$, times an exponential cutoff, with cutoff energy $110 \pm 50$ GeV~\cite{Fermi-LAT:2014sfa}. We assume the same modeling for the astrophysical component:
\begin{eqnarray}
\Phi_{\rm astrophysical}=N\  {E}^{-\alpha} e^{E/E_{\rm cut}} .
\label{F-B}
\end{eqnarray}
We leave $\alpha$ as a free parameter to test if we can recover the known {\it Fermi}-bubble spectral index above $10^{\circ}$ in Galactic latitude.

Regarding the DM part, we assume the $S$ particles of the ESHP model acting as DM, and compute the corresponding emission spectrum  in the ESHP parameter space, $\{\lambda_S, \lambda', m_S\}$ using MicrOmegas 4.3.2 \cite{Belanger:2014vza}. Since the main annihilation channel is the one depicted in Fig.~ \ref{fig:processes} (right panel), i.e. $SS\rightarrow h h$, most of the photons come from the subsequent decay of the Higgs-bosons into $b\bar b$, but there are other contributions coming from $h\rightarrow WW$ and even $h\rightarrow \gamma\gamma$ (the latter gives an interesting spectral feature, as we will see below). The prompt Galactic Center flux coming from DM annihilations, $\Phi_{\rm DM}$, is proportional to this spectrum times the so-called $J-$factor
\begin{eqnarray}
J_{10^{0}}=\int_{\Delta\Omega} d\Omega\int_{\rm l.o.s.} ds \rho^2(r(s,\theta))\ ,
\label{J}
\end{eqnarray}
where $\rho$ is the DM density, $r$ is the spherical distance from the Galactic Center, $\theta$ is the observational angle towards the GC and $s$ is the line of sight (l.o.s.) variable. As usual, we assume a NFW profile
\begin{eqnarray}
\rho(r) = \rho_s \left(\frac{r}{r_s}\right)^{-\gamma} \left(1+\frac{r}{r_s}\right)^{-3+\gamma} \ ,
\label{NFW}
\end{eqnarray}
where $r_s$ is the scale radius (20 kpc), $\rho_s$ a scale density fixed by requiring the local DM density at the 8.5 kpc Galactocentric radius to be 0.4 GeV cm$^{-3}$; and $\gamma=1.25\pm0.8$, as given in ref.~\cite{1704.03910}. 

In summary, the fit contains 6 independent parameters: $\{N, \alpha, E_{\rm cut}\}$ (astrophysical part) and  $\{\lambda_S, \lambda', m_S\}$ (DM part).  In order to assess the quality of a fit we construct the $\chi^2-$function:
\begin{eqnarray}
\chi^2=\sum_{i,j} (\Phi_i^{\rm obs}-\Phi_i^{\rm m})\hat\Sigma_{i,j}^{-1}(\Phi_j^{\rm obs}-\Phi_j^{\rm m}).
\label{chi2}
\end{eqnarray}
Here $i$ labels the energy-bin, $\Phi_i^{\rm m}$ is the flux for a model ($m$), defined by the values of the above six parameters, $ \Phi_i^{\rm obs}$ is the derived flux with the Sample model, the light blue points in Fig.~\ref{fig:GCE}, and $\hat\Sigma_{i,j}^{-1}$ is the inverse of the covariance matrix, which was derived in ref.~\cite{Caron:2017udl}. Note that the derived information on the GCE spectrum in ref.~\cite{1704.03910} is enclosed in $\{\Phi_i^{\rm obs},  \hat\Sigma_{i,j}^{-1}\}$. 
Since the functions used to fit the GCE are not linear, we cannot use the reduced $\chi^2$ to compute $p-$values\footnote{For a detailed discussion see ref~\cite{2010arXiv1012.3754A}.}. Instead, following ref.~\cite{Caron:2017udl}, we will proceed in this way:

\begin{enumerate}

\item For each point under consideration in the ESHP parameter-space (defined by $\{\lambda_S, \lambda', m_S\}$), we allow the astrophysical parameters, $\{N, \alpha, E_{\rm cut}\}$ to vary, in order to find the best fit to the data. 
This gives $\Phi^{\rm m}_{\rm best}$.

\item Create a set of $10^7$ pseudo-random (mock) data normal-distributed with mean at $\Phi^{\rm m}_{\rm best}$, according to $\hat\Sigma_{i,j}$

\item Compute $\chi^2$ for each data created in step 2.

\item Create a $\chi^2$ distribution using the values from step 3.

\item The integrated $\chi^2$ distribution up to the best-fit-$\chi^2$ to the actual data, gives the $p-$value of the model.

\end{enumerate}

It turns out that the shape of the $\chi^2$ distribution is extraordinarily stable through the whole parameter space, so it can be settled once and for all, with a consequent saving of computation time. The $\chi^2$ distribution is illustrated in Fig.~\ref{fig:X2distr} below.

In addition to the fit of the GCE data, we require that every point is not constrained by other DM detection observables like the spin-independent cross-section from the XENON1T experiment \cite{Aprile:2017iyp} and the thermal averaged annihilation cross section from the search of DM  in dwarf spheroidal satellite galaxies of the Milky Way (dSphs) by the {\it Fermi}-LAT experiment \cite{Ackermann:2015zua}.

The spin-independent cross section is evaluated analytically. Parametrizing the Higgs-nucleon coupling as $f_N m_N /v$, where $m_N\simeq 0.946$ GeV is the mass of the nucleon, the spin-independent cross section, $\sigsi$, reads
\begin{equation}
\sigsi
\label{sics}=
\frac{\lambda_1^2 f_N^2 \mu^2 m_N^2}{4 \pi m_h^4 m_{S}^2}\ ,
\end{equation}
where $\mu=m_N m_{S}/(m_N+m_{S})$ is the nucleon-DM reduced mass. The $f_N$ parameter contains the nucleon matrix elements,
and its full expression can be found, e.g., in ref.~\cite{Cline:2013gha}.
Using the values for the latter obtained from the lattice evaluation
\cite{Alarcon:2011zs,Alarcon:2012nr,Alvarez-Ruso:2013fza,Young:2013nn,Abdel-Rehim:2016won,Duerr:2016tmh}, one arrives
at $f_N=0.30 \pm 0.03$, in agreement with Ref.\,\cite{Cline:2013gha}. The strongest results on spin-independent cross section are given by XENON1T.

Concerning constraints from dSphs, we use gamLike 1.1 \cite{Workgroup:2017lvb}, a package designed for the evaluation of likelihoods for $\gamma$-ray searches which is based on the combined analysis of 15 dSphs using 6 years of {\it Fermi}-LAT data, processed with the pass-8 event-level analysis. For any point in the parameter space we scale the photon flux by the $\xi^2$ factor, with $\xi\equiv \Omega_S/\Omega_{CDM}$. GamLike  provides a combined likelihood, with which we perform the test statistic \cite{Rolke:2004mj}
\begin{equation}
TS=-2 \ln( \mathcal{L}(\mu, \theta  \mid D)/\mathcal{L}(\mu_0, \theta \mid D)) \; ,
\end{equation}
where $\mu$ denotes the parameters of the DM model, $\mu_0$ corresponds to no-annihilating DM, $\theta$ are the nuisance parameters used in the {\it Fermi}-LAT analysis \cite{Ackermann:2015zua}, and  $D$ is the $\gamma$-ray data set .
To find a 90\% upper limit on the DM annihilation cross-section we look for changes in $TS = 2.706$.

\section{Results}

\begin{figure}[h!]
\centering 
\includegraphics[width=0.8\linewidth]{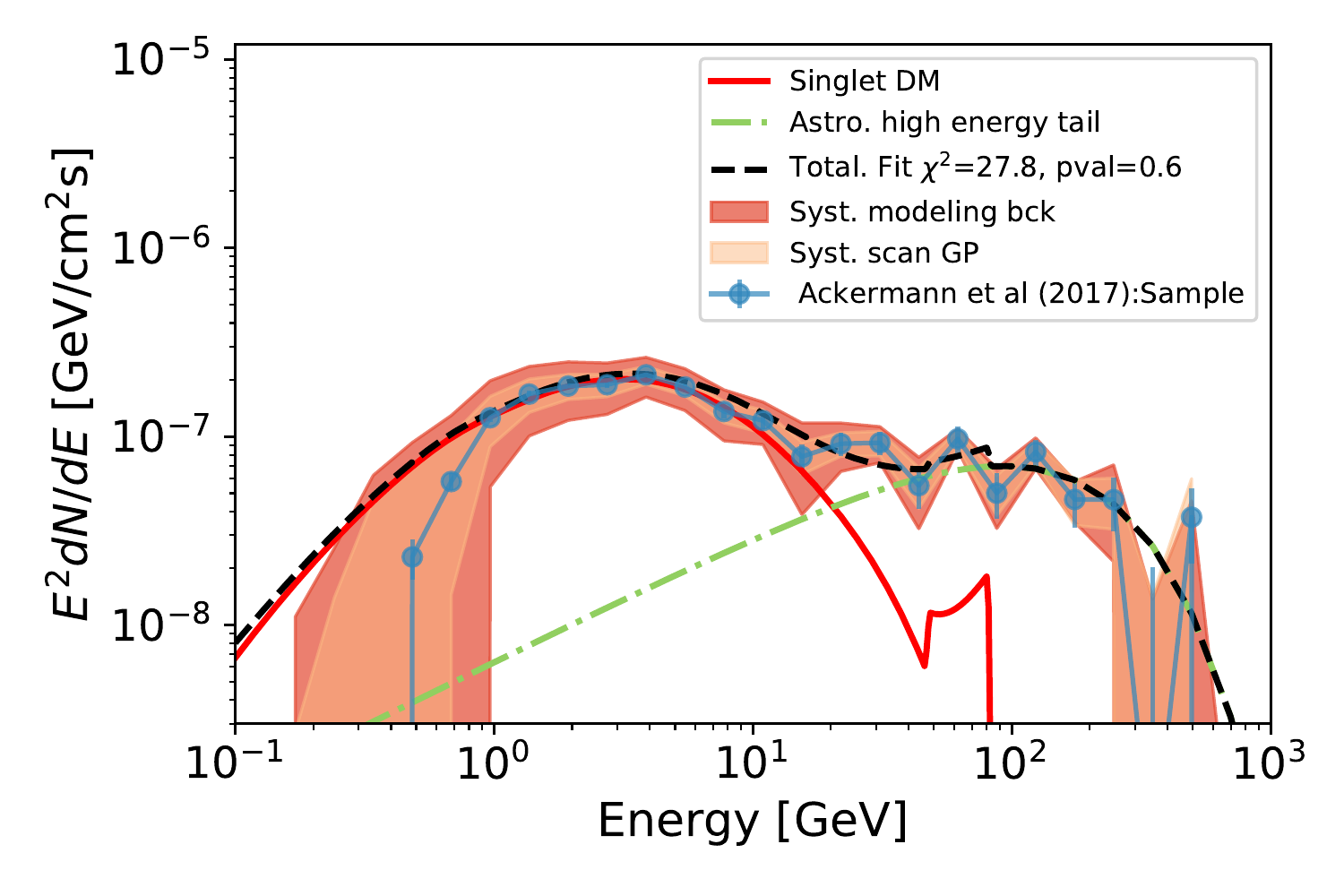}
\caption{
Fit to the GCE spectrum (blue dots of Fig.~\ref{fig:GCE}) by the combination of a power-law  with an exponential cutoff, describing the astrophysical sources (green dash-dotted line), plus the contribution of DM annihilation, as given by the ESHP model (red line); with parameters given in eq.(\ref{fit}). See eqs. (\ref{flux}-\ref{NFW}) for further details. The black dashed line gives the final prediction of the model.
}
\label{fig:fit}
\end{figure}

\begin{figure}[h!]
\centering 
\includegraphics[width=0.6\linewidth]{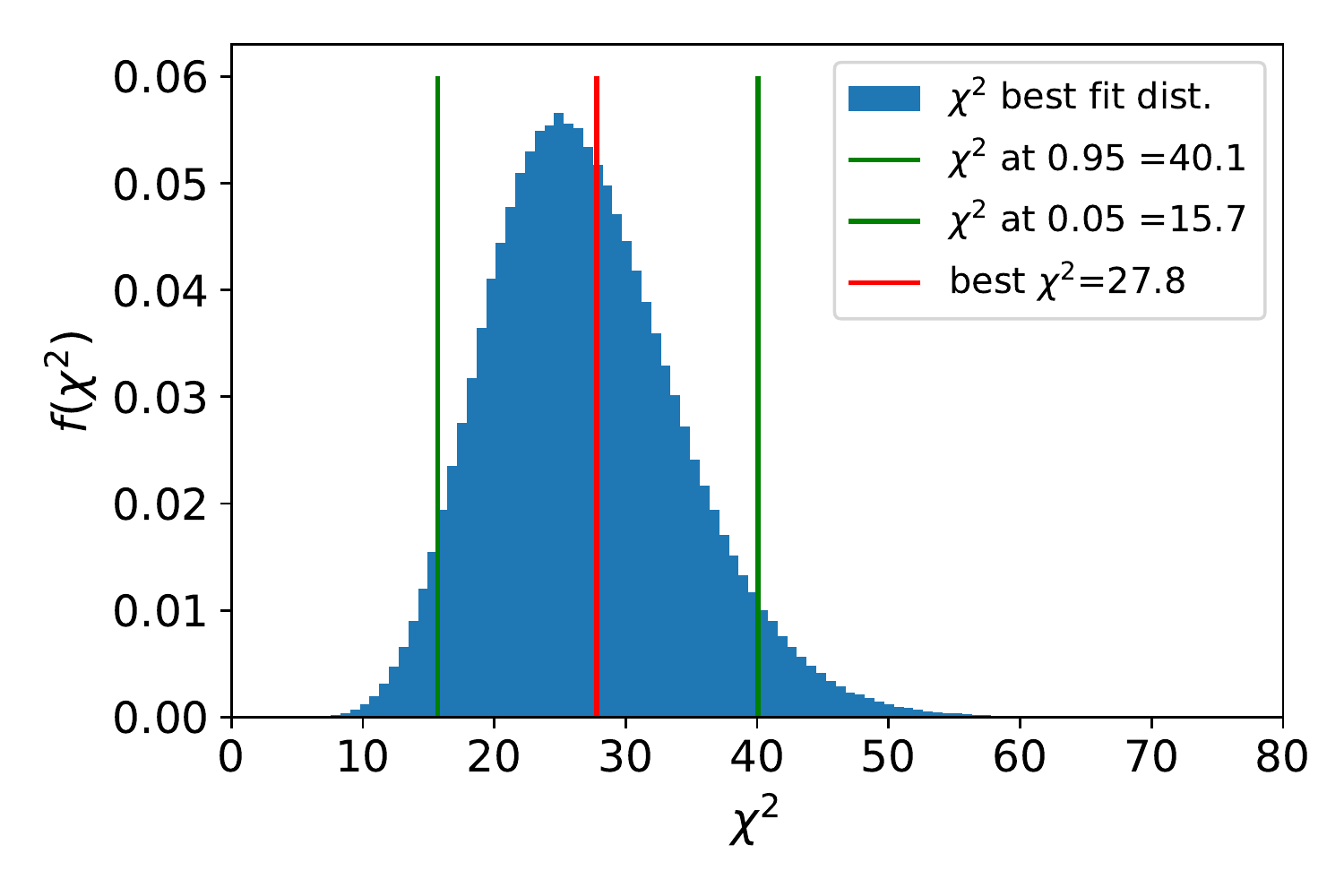}
\caption{
The blue histogram represents the distribution of $\chi^2$ drawn from the best fit to the GCE spectrum in Fig. \ref{fig:fit}, green vertical lines correspond to upper and lowers limits at 5\%. The red vertical line represents the best fit model in Fig.\ref{fig:fit}.
}
\label{fig:X2distr}
\end{figure}

Despite having just three parameters,  $\{\lambda_S, \lambda', m_S\}$, the ESHP model has regions of the parameter space that could be contributing significantly to the GCE without conflicting with other observables. In fact, this holds even if one of the parameters, $\lambda_S$ (the initial $S^2|H|^2$ coupling in the ordinary SHP model) is set to zero, since the $\lambda'-$coupling is enough to lead to sufficient DM annihilation to reproduce the correct relic density, without conflicting with direct-detection, as explained in section 2. Then, the ESHP model may work with just two parameters, as the standard SHP. A representative point in the ESHP parameter space, not rejected as a possible explanation of a significant fraction of the GCE is illustrated in Fig.~\ref{fig:fit}, which corresponds to the following values of the parameters:
\begin{eqnarray}
&&m_S= 131\ {\rm GeV},\;\;\; \lambda_S = 0,\;\;\; \lambda' = 0.58,
\nonumber\\
&& \alpha=1.5,\;\;\; E_{\rm cut}= 178\ {\rm GeV}\ .
\label{fit}
\end{eqnarray}
Note that the $\lambda_S$ coupling is set to zero, so the value of $\lambda'$ is simply the required one to reproduce the correct relic density.
The astrophysical exponent, $\alpha$, becomes close to the estimations from the {\it Fermi}-buble emission at high latitudes, $\alpha\simeq 1.9$.\footnote{Fixing $\alpha$ in the fitting procedure at the value preferred by the {\it Fermi}-bubble analysis, $\alpha = 1.9$, is also possible. Then typically, less flux from DM annihilation is required at low-energy and, consequently, the favoured values of the
 $\lambda'-$coupling are somewhat smaller.} The fit is quite good, with $p-$value=0.63 (coresponding to a $\chi^2= 27.8$ for the 27 energy-bins). 
As mentioned in the previous section, this $p-$value is obtained from the associated $\chi^2$ distribution, which for this particular point is shown in Fig.~\ref{fig:X2distr}.

Fig.~\ref{fig:fit} also shows an amusing peculiarity. Namely, the  $h\rightarrow \gamma\gamma$ decay contributes very little to the total flux, but located around a typical energy $E\simeq m_S/2 \simeq 65$ GeV. This corresponds to a visible feature in the red line, which produces a bump in the total flux in a bin where data show a peak as well. The feature, however, is spread over the $\sim 40-80$  GeV range
since the Higgs giving the two photons has a non-vanishing momentum. 
Consequently, the usual {\it Fermi}-LAT contraints on $\gamma$-lines are not applicable here. In addition, the total flux coming from this process is below the present limits on lines at $\sim 65$ GeV \cite{Ackermann:2015lka}. So, even if it were concentrated at that energy it would be non-detectable yet. Nevertheless, it is not unthinkable that a future dedicated search could be sensitive to this feature.

%

\begin{figure}[h!]
\centering 
\includegraphics[width=0.6\linewidth]{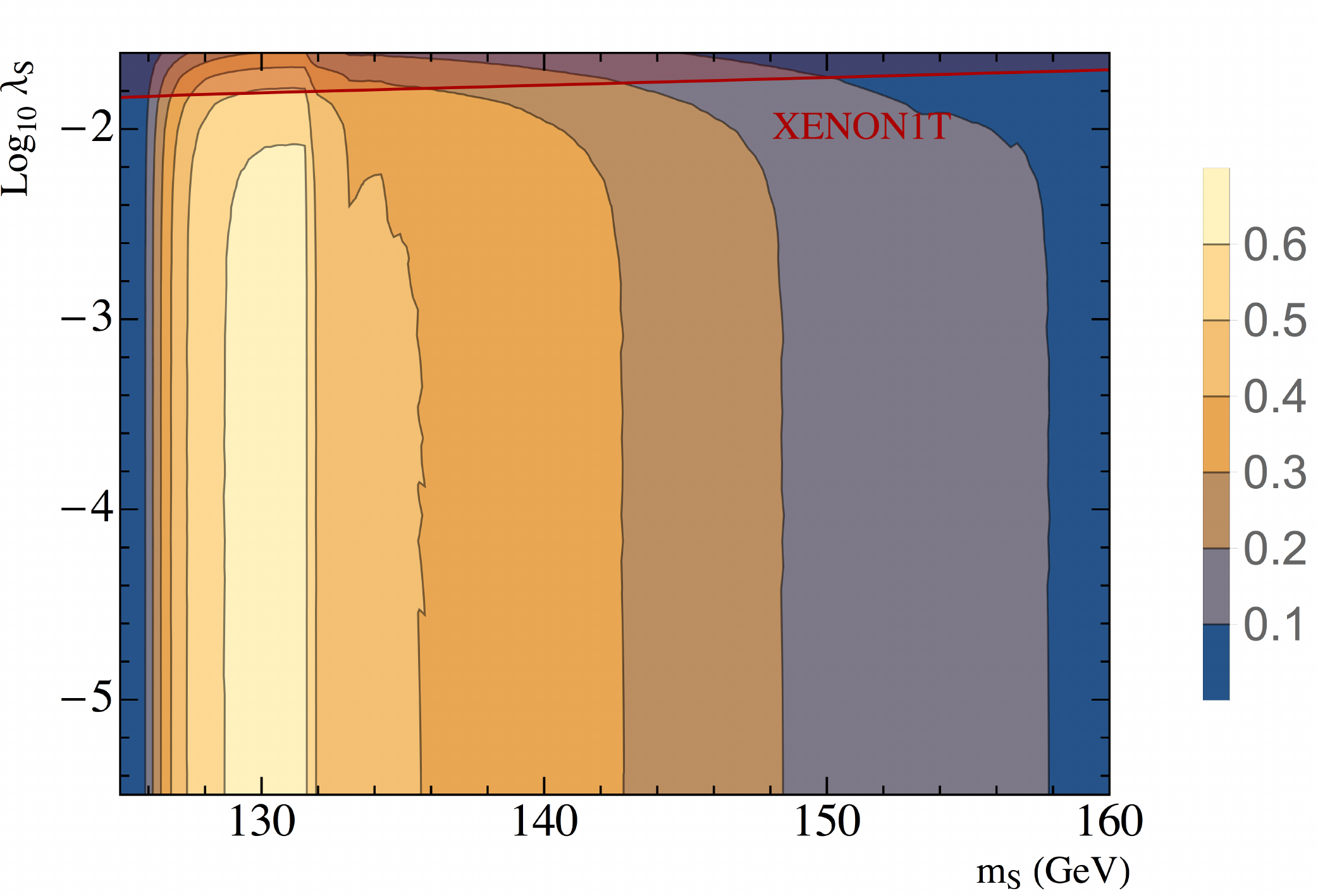}
\caption{
{Contours of constant $p-$value in the $\{m_S, \lambda_S\}$  plane in the context of the ESHP model. The value of $\lambda'$ is adjusted at each point in order to reproduce the correct relic density. The XENON1T direct-detection bound is shown (red line).  For small $\lambda_S$ the plot lacks structure in the vertical axis.}
}
\label{fig:Omegacorrecta}
\end{figure}

Fig.~\ref{fig:Omegacorrecta} shows the $p-$value in the $\{m_S, \lambda_S\}$ plane, where $\lambda'$ is adjusted for each point in order to reproduce the correct relic density. The XENON1T bound is also shown. As expected, it only gives restrictions when $\lambda_S$ is sizeable, which is not necessary. The constraints from dSphs do not appear, as they do not give any constraint. Obviously, for small $\lambda_S$ the plot lacks structure in the vertical axis.

\begin{figure}[h!]
\centering 
\includegraphics[width=0.43\linewidth]{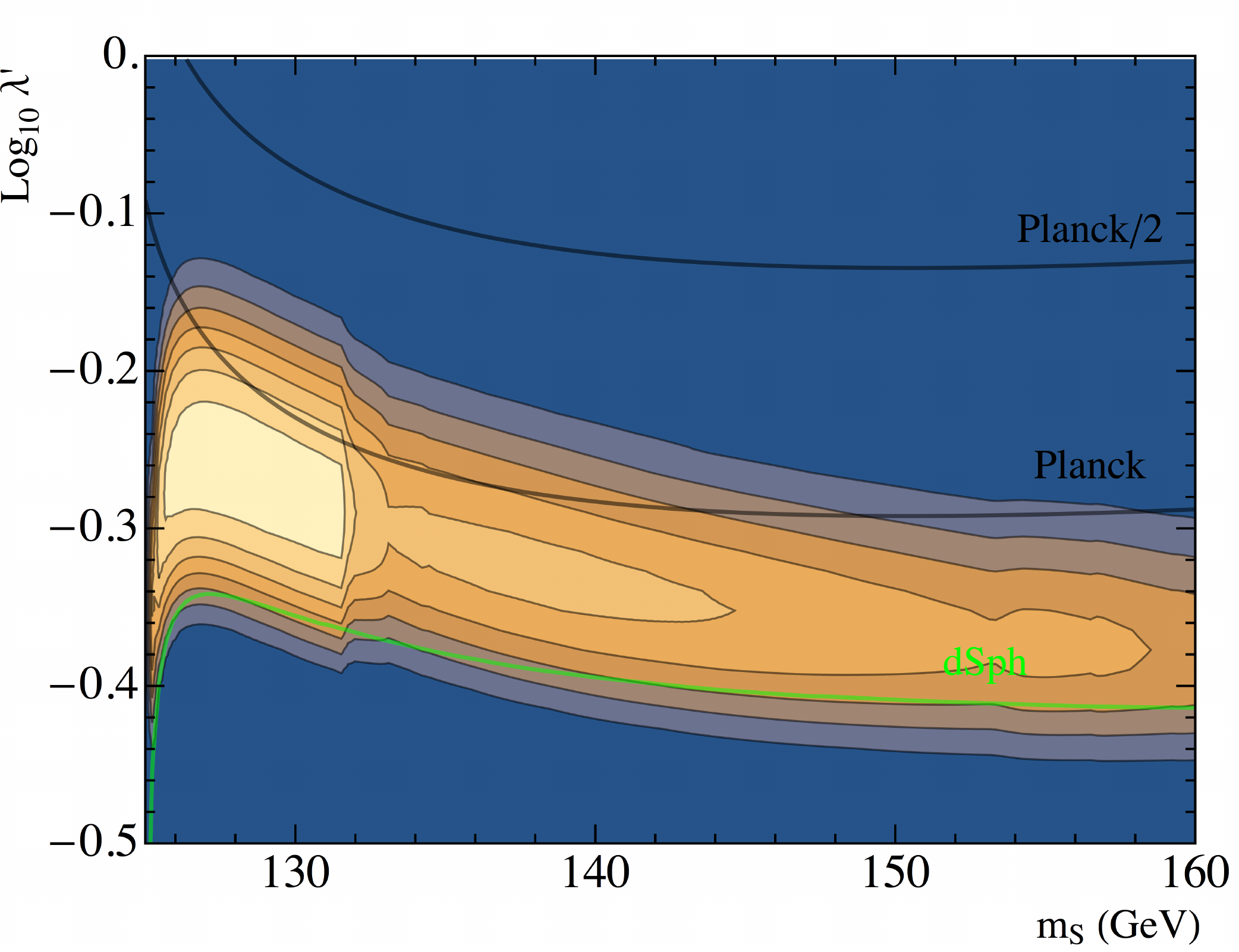}
\includegraphics[width=0.52\linewidth]{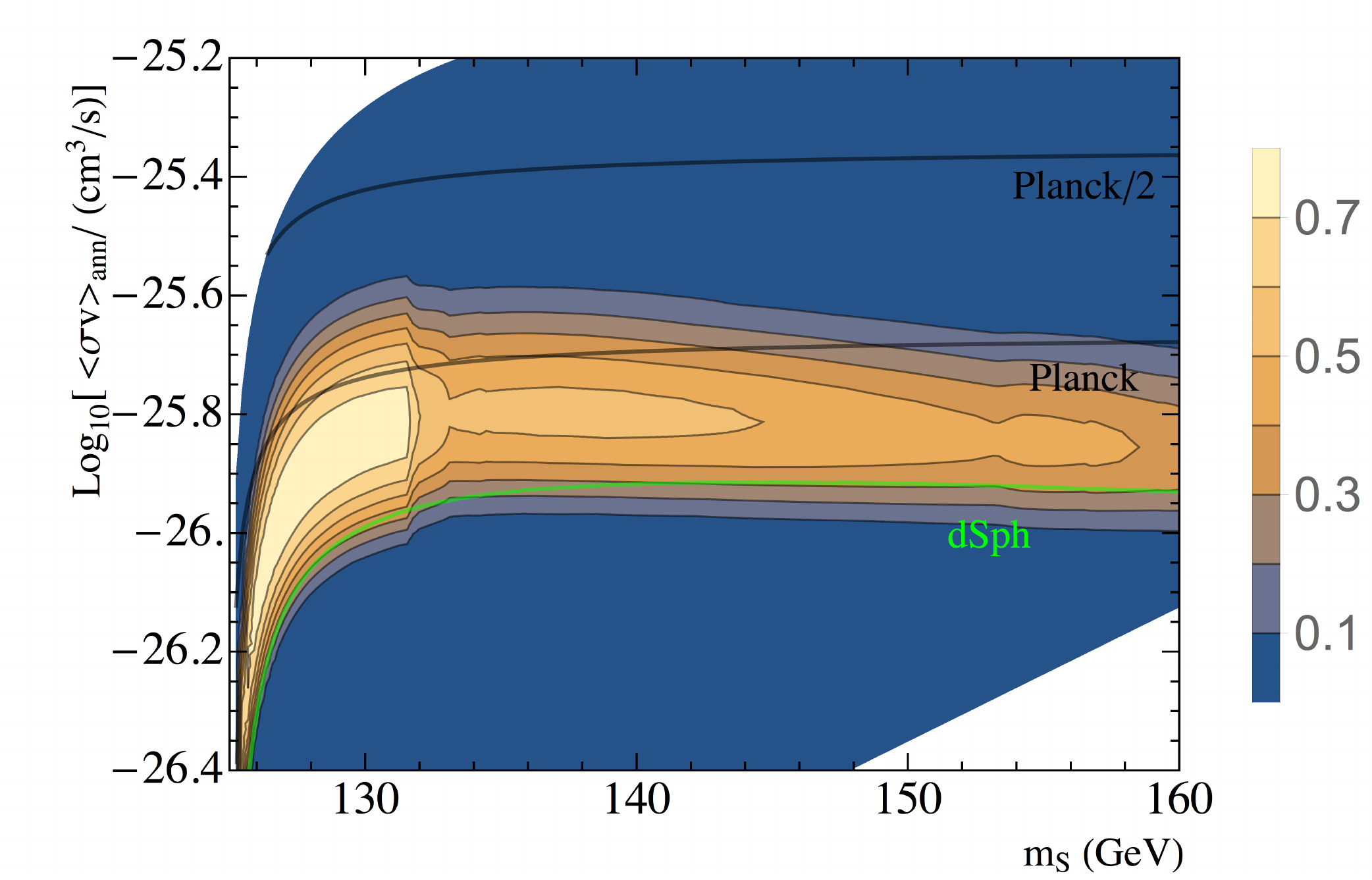}
\caption{
Contours of constant $p-$value in the $\{m_S, \lambda'\}$ plane (left panel) and the $\{m_S,  \langle\sigma_{\rm ann} v\rangle\}$ 
plane (right panel), setting $\lambda_S=0$. Now, the value of the relic density depends on the point, increasing in the downward direction.
The lower (upper) black curve corresponds to (half of) the Planck relic-density. The green curve shows the lower bound on $\lambda'$ from dwarf spheroidal observations. 
}
\label{fig:lambda_1}
\end{figure}

Fig.~\ref{fig:lambda_1}  (left panel)  is an equivalent plot where $\lambda_S$ has been set to zero, so that $\{m_S, \lambda'\}$ are the only relevant parameters. Now, the value of the relic density depends on the point. The lower black curve corresponds to the Planck relic-density, hence it coincides with the horizontal bottom line of Fig.~\ref{fig:Omegacorrecta}. Below that curve the relic density is too high. The upper black curve corresponds to half of the relic density. Interestingly, models in the parameter space that reproduce the whole dark matter relic density with $S$ particles are the ones with higher $p-$values. Indeed, the regions with an optimal fit of the GCE present a (slightly) too-large relic density, implying that the points along the "Planck"-line tend to produce (slightly) less GC flux than required (recall here that the annihilation cross section of dark matter increases as $(\lambda')^2$, while the J-factor goes as $\rho_{\rm DM}^2\sim (\lambda')^{-4}$ ). In consequence, the possibility commented in the Introduction that only a fraction of the low-energy GCE is associated to DM annihilation is still the most advantageous one. However, in this scenario that fraction is remarkably close to the whole GCE.

The green curve gives the lower bound on $\lambda'$ from dwarf observations, taking into account the previosly mentioned $\xi^2$ factor. 
Clearly, dSphs limits do not impose any constraint in practice. Actually, it is apparent that the green curve corresponds to $\xi>1$. Therefore, assuming that in that region of the parameter space the relic density is the observed one (thanks to some unspecified mechanism), as it is sometimes done, then the dSphs limit becomes even weaker.

Fig.~\ref{fig:lambda_1} (right panel) is the equivalent plot in the $\{m_S, \langle\sigma_{\rm ann} v\rangle\}$ plane.

\section{Conclusions}

The {\it Fermi}-LAT Collaboration has recently presented a new analysis of the Galactic Center area based on the reprocessed Pass 8 event data, giving insight on the GCE emission. The analysis confirms the presence of the GCE, now peaked at the $\sim 3$ GeV region, i.e. slightly shifted towards higher energies.

The GCE could be originated by the sum of a {\it Fermi}-bubble like emission plus another source, which might be DM annihilation, unresolved MSP, or a combination of both. Certainly, the interpretation of the GCE as DM emission is controversial but it remains an interesting possibility. 
On the other hand, such instance is not easy to implement in a particular, theoretically sound, DM model. Part of the difficulty comes from the fact that a DM model able to reproduce the GCE typically leads also to direct detection predictions  which are already excluded by XENON1T. This is the case for one of the most economical and popular scenarios of DM, namely the Scalar Singlet-Higgs-Portal (SHP), where the DM particle is a singlet scalar, $S$, that interacts with the SM matter through couplings with the Higgs  \cite{Cuoco:2016jqt}.

In this paper we have considered a particularly economical extension of the SHP, the so-called ESHP, which simply consists in the addition of a second (heavier) scalar singlet in the dark sector. This extra particle can be integrated-out, leaving a standard SHP plus a dimension-6 operator, $\propto S^2(|H|^2-v^2/2)^2$, which reproduces very well the ESHP results. Hence, the model just adds one extra relevant parameter to the ordinary SHP. Actually, the usual DM-Higgs quartic coupling of the SHP, $\lambda_S S^2|H|^2$, can be safely set to zero (or a negligible value) since the dimension-6 operator can take care of all the phenomenology. The main virtue of the ESHP is that it allows to rescue large regions of the SHP parameter-space, leading to the correct relic density and avoiding the strong direct-detection constraints, as it does not imply any effective trilinear $S^2h$ coupling. Concerning DM annihilation, the main channel is into two Higgs bosons, $SS\rightarrow h h$; thus most of the photons in the final state come from the subsequent decay of the latter into $b\bar b$.

We have shown that, in large regions of the parameter space, the ESHP model produces excellent fits to the GCE in the $E= 1-10$ GeV region. The region above $10$ GeV is well described by an additional  power-law component which accounts for astrophysical {\em Fermi}-bubble-like contributions.
Those favoured regions of the ESPH parameter-space are not in conflict with other observables, and reproduce the DM relic density at the Planck value. This is illustrated in Fig.~\ref{fig:fit}, which shows a very good fit, with $p-$value=0.63 (corresponding to a $\chi^2= 27.8$ for the 27 energy-bins), obtained with representative values of the model parameters, in particular the DM particle candidate has a mass $m_S\simeq 130$ GeV. The secondary annihilation channel $h\rightarrow \gamma\gamma$, which contributes very little to the total flux, concentrates around a typical energy $E\simeq m_S/2 \simeq 65$ GeV. This produces a bump in the total flux in a bin where data show a peak as well.

We have demonstrated the global performance of the model by scanning the parameter space, both demanding correct relic density and allowing for a smaller one (which would require other DM components). Figs.~\ref{fig:Omegacorrecta}, \ref{fig:lambda_1} show that large portions of the parameter space provide a sensible description of the GCE. We have also checked that direct-detection (XENON1T) and dSphs observations do not impose any relevant constraints in practice.

In summary, the ESHP model, which is a very economical extension of the popular scalar-singlet Higgs portal (SHP) model, 
provides a fair description of the GCE for very reasonable values of the parameters (unlike the SHP), in particular $m_S \gsim 130$ GeV, while keeping the correct DM relic density, and without conflicting with other direct and indirect detection data.

\section*{Acknowledgements}

This work has been partially supported by MINECO, Spain, under contracts FPA2014-57816-P, FPA2016-78022-P  and Centro de excelencia Severo Ochoa Program under grants SEV-2014-0398 and SEV-2016-0597.  The work of GAGV was supported by Programa FONDECYT Postdoctorado under grant 3160153. The work of J.Q. is supported through the Spanish FPI grant SVP-2014-068899. R. RdA is also supported by the Ram\'on y Cajal program of the Spanish MICINN, the Elusives European ITN project (H2020-MSCA-ITN-2015//674896-ELUSIVES) and the  ``SOM Sabor y origen de la Materia" (PROMETEOII/2014/050). We thank the Galician Supercomputing Center (CESGA) for allowing the access to Finis Terrae II Supercomputer resources.

\providecommand{\href}[2]{#2}\begingroup\raggedright\endgroup

\end{document}